\def\gs{\mathrel{\hbox{\rlap{\hbox{\lower4pt\hbox{$\sim$}}}\hbox{$>$}}}}
\def\ls{\mathrel{\hbox{\rlap{\hbox{\lower4pt\hbox{$\sim$}}}\hbox{$<$}}}}

\documentclass[useAMS,usegraphicx]{mn2e}
\usepackage{rotate}
\usepackage{times}
\newif\ifAMStwofonts
\AMStwofontstrue

\title[{\em XMM-Newton} spectral properties of the  Narrow-Line
Seyfert~1 galaxy IRAS 13224--3809]
{{\em XMM-Newton} spectral properties of the
Narrow-Line Seyfert 1 galaxy IRAS 13224--3809 }

\author[Th. Boller et al.]
{Th. Boller,$^1$ Y. Tanaka,$^{1,2}$ A. Fabian,$^3$ W. N. Brandt,$^4$ L. Gallo,$^1$ N. Anabuki,$^2$
\newauthor Y. Haba$^2$ and S. Vaughan$^3$ \\
$^1$ Max-Planck-Institut f\"ur extraterrestrische Physik, Postfach 1312, 85741 Garching, Germany \\
$^2$ Institut of Space and Astronautical Science, 3-1-1 Yoshinodai, Sagamihara, Kanagawa 22, Japan\\
$^3$ Institute of  Astronomy, Madingley Road, Cambridge CB3 0HA\\
$^4$ Department of Astronomy and Astrophysics, Pennsylvania State University, 525 Davey Lab, University Park, PA 16802, USA
}

\date{Received 21 March 2003 ; Accepted 13 May 2003}

\begin{document}
\label{firstpage}
\maketitle

\begin{abstract}
We report on the first {\em XMM-Newton} observation of the highly
X-ray variable, radio-quiet, Narrow-Line Seyfert 1 galaxy IRAS
13224--3809 obtained during the guaranteed time programme with a
64 ks exposure. The most remarkable spectral feature is a sharp
drop, by a factor $\sim 5$, in the spectrum at 8~keV. This is a
similar, but stronger, feature to that which we found in
1H\,0707--495. Significant flattening of the hard X-ray spectrum
occurs when the source flux decreases. The flattening of the
spectrum can be modelled as an increase in the column density of
the absorbing material and/or its covering fraction. Below
$\sim1.5$~keV the spectrum is dominated by a giant soft X-ray
excess, and at around $\sim 1.2$~keV there is a significant
absorption feature detected, most probably due to ionized Fe L
absorption. The new X-ray spectral properties detected with {\em
XMM-Newton} in IRAS 13224--3809 support a partial covering
interpretation, i.e. the presence of dense material
inside the accretion-disc region partially obscuring the emission
from the accretion disc. However, the sharpness of the feature, if
due to photoelectric absorption, is surprising and may require an
alternative explanation. One possibility which does fit the whole
spectrum is that it is dominated by ionized reflection rather than
absorption. The unusual spectral properties detected with {\em
XMM-Newton} from Narrow-Line Seyfert 1 galaxies increase the known
spectral complexity of active galactic nuclei and should further
stimulate a combined theoretical and observational effort to
achieve a better understanding of the physics of the innermost
regions of AGN.
\end{abstract}


\section{Introduction}

{\em ROSAT}, {\em ASCA} and {\em XMM-Newton} have shown many
Narrow-Line Seyfert~1 galaxies (hereafter NLS1s; see Osterbrock \&
Pogge 1985 and Goodrich 1989) to have remarkable X-ray properties
compared to Seyfert~1 galaxies with broader Balmer lines. In X-rays
NLS1s are generally characterized by strong soft X-ray excesses, steep
hard power-law continua and extreme X-ray variability
(e.g. Puchnarewicz et al. 1992, Boller, Brandt \& Fink 1996, Brandt \&
Boller 1998, Vaughan et al. 1999,  Brandt, Mathur \& Elvis
1997). With the arrival of {\em XMM-Newton} spectra a new feature was added:
a sharp spectral drop above 7 keV
without any significant Fe-K line emission in the NLS1 1H~0707--495 (Boller et al. 2002; Fabian et
al. 2002).   In the case of 1H~0707--495 two
physical interpretations were discussed: (i) a partial covering
scenario (following Rees 1987; Celotti, Fabian \& Rees 1992; Brandt \&
Gallagher 2000) in which the sharp drop is interpreted as
Fe absorption in high column density clouds (Boller et al. 2002) and (ii) a reflection-dominated accretion-disc spectrum
(Fabian et al. 2002). Both models provide
acceptable spectral fits to the 1H~0707--495 data. The very high iron
overabundance in the Boller et al. (2002) interpretation was reduced
to a more realistic value by the reanalysis of Tanaka et al.  (2003).

IRAS 13224--3809 is one of the most interesting members of the NLS1
class due to its remarkable X-ray variability, huge soft X-ray excess
(Boller, Brandt \& Fink 1996) and high optical Fe II to H$\rm \beta$
line ratio (Boller et al. 1993). Indeed, IRAS 13224--3809 is among the
most X-ray variable Seyfert 1 galaxies known. The first systematic
monitoring in soft X-rays in 1996 showed persistent, rapid,
giant-amplitude count rate variability. Over the course of the
observations the maximum observed amplitude of variability was of a
factor of $\sim 60$, and a variation by about a factor of 57 in just
two days was observed. The ionizing luminosity rises from about $1.5
\times 10^{43}$ erg s$^{-1}$ to about $8.3 \times 10^{44}$ erg s$^{-1}$,
roughly equivalent to a typical Seyfert 1 like
MCG--6-30-15 abruptly increasing its soft X-ray luminosity to become
almost as powerful as a quasar (Boller et al 1997).

Due to its extreme X-ray spectral and timing properties, IRAS
13224--3809 was observed during the {\em XMM-Newton} guaranteed time
programme. In this paper we discuss the results obtained from the
64~ks observation.

\section{X-ray observations and data analysis}

\begin{figure}
 \includegraphics[width=6 cm, angle=270]{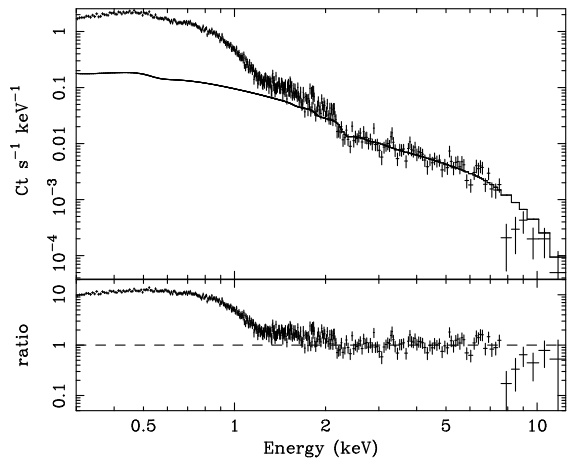}
 \caption{EPIC pn spectrum of IRAS 13224--3809.
A simple power-law model, fitted only in the 2--7 keV range, has been extrapolated
into the soft and hard bands to illustrate the sharp spectral feature
at $\sim$8~keV, and  the  soft X-ray excess emission. The photon index
in the 2--7 keV band is (1.9 $\rm \pm$ 0.1).}
\end{figure}

IRAS 13224--3809 was observed with {\em XMM-Newton} (Jansen et
al. 2001) on 2002 January 19 during revolution 0387 for about 64 ks.
During this time all instruments were functioning normally.  The EPIC
pn camera (Str\"uder et al. 2001) was operated in full-frame mode, and
the two MOS cameras (MOS1 and MOS2; Turner et al.  2001) were operated
in large-window mode.  All of the EPIC cameras used the medium filter.
The two Reflection Grating Spectrometers (RGS1 and RGS2; den Herder et
al. 2001)
also gathered data during this time. The Observation Data Files (ODFs)
were processed to produce calibrated event lists using the {\em
XMM-Newton} Science Analysis System ({\tt SAS v5.3}). Unwanted hot,
dead or flickering pixels were removed as were events due to
electronic noise.  Event energies were corrected for charge-transfer
losses.  The latest available calibration files were used in the
processing.  Light curves were extracted from these event lists to
search for periods of high background flaring. A significant
background flare was detected in the EPIC cameras approximately 20 ks
into the observation and lasted for ~5 ks. This segment was removed
and ignored during the analysis. The total amount of good exposure
time selected was 55898 s and 58425 s for the pn and MOS detectors,
respectively. The source plus background photons were extracted from a
circular region with a radius of 35 arcsec, and the background was
selected from an off-source region and  appropriately scaled to the source
region.  Single and double events were selected for the pn
detector, and single-quadruple events were selected for the MOS. High
resolution spectra were obtained with the RGS.  The RGS were operated
in standard Spectro+Q mode for a total exposure time of 63963 s.  The
first-order RGS spectra were extracted using RGSPROC, and the response
matrices were generated using RGSRMFGEN.
The Optical Monitor collected data through the UVW2 filter
(1800--2250
\AA) for about the first 25 ks of the observation.    The
OM data are discussed in Gallo et al. (2003,  in preparation).
In the following analysis we use the EPIC pn data  to constrain the sharp spectral feature at
8~keV, as they contain the highest photon statistics.  The energy
range up to about 12~keV can be explored with EPIC pn while above that
value the spectrum is background dominated.  The combined MOS1 and
MOS2 spectrum is affected by high background at energies greater than
6~keV. Above 8 keV, we collect a total of 120 source plus background counts, while
the number of background photons, normalized to the source cell size, is 68.

\section{X-ray spectroscopy}

\subsection{Mean spectral properties}

\subsubsection{Discovery of a sharp spectral feature}

\begin{figure}
 \includegraphics[width=7.5 cm, angle=270]{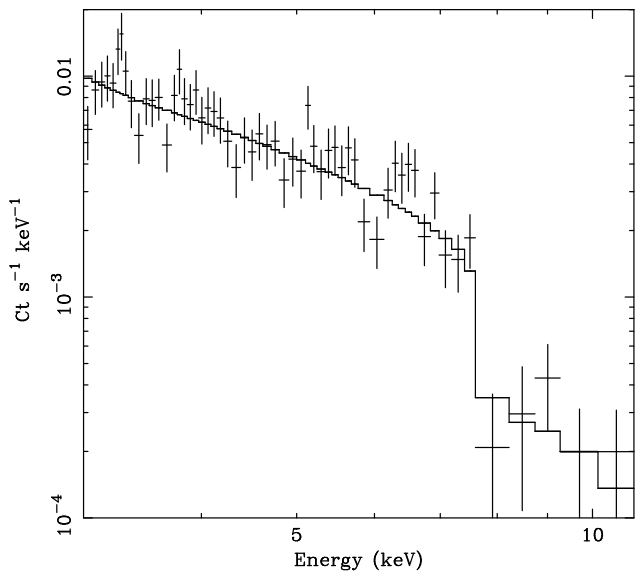}
 \caption{Simple power-law plus edge fit to the 2--12~keV energy
 range. The edge parameters are $\tau = 1.6_{-0.5}^{+0.9}$ and $E = 8.2\pm0.1$~keV.
 }
\end{figure}

In Fig.~1 we show the time-averaged spectrum of IRAS 13224--3809. We
have fitted the 2--7 keV energy band with a simple power-law,
which is extrapolated into the soft and hard bands for illustration
purposes. The most obvious features are the strong soft X-ray
excess emission below about 1.5 keV and the presence of a sharp spectral
drop at $\sim 8$~keV.
If the 2--12 keV spectrum is fitted by a simple power-law plus edge model,
the edge energy is $\rm 8.2 \pm 0.1 $ keV (Fig.~2).
The energy of the edge is significantly different from that of the
neutral iron absorption edge at 7.1~keV. The flux absorbed by the edge
is $4.6 \times 10^{-14}$ erg cm$^{-2}$ s$^{-1}$. The edge cannot be
constrained by the MOS data as the spectrum is background dominated above
about 6 keV.

Some positive residuals can be seen at around 6.8~keV (in the rest
frame of the object). They are consistent with emission from H and
He-like iron with a total equivalent width of about 200~eV. The
detection of this emission, judged from an $F$-test, is marginal (at
about the $2.5\sigma$ confidence level). The strength of the line,
when compared with the flux of photons which is absorbed and the
likely fluorescent yield (Krolik \& Kallman 1987), indicates that the
absorber covers a solid angle $\Omega/4\pi\ls 0.2$.

\subsubsection{A partial covering model}

\begin{figure}
 \includegraphics[width=5.5 cm, angle=270]{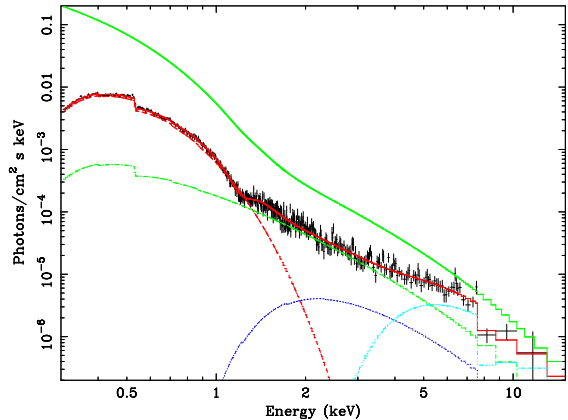}
 \caption{Unfolded spectrum modelled
with a partially covered power-law and a disc blackbody. The upper curve gives the
absorption-corrected spectral model (see text for details).}
\end{figure}

The observed spectrum shows little low-energy absorption and is
consistent with interstellar absorption only. On the other hand, there
is what appears to be a deep absorption edge at $8.1-8.2$~keV which
would correspond to the K-edge of Fe~{\sc xviii--xx}. The sharpness of
the edge requires that the ionization states of Fe are confined to a
narrow range. The absence of strong (narrow) emission at $\sim6.6$ keV
is in contradiction with an explanation in terms of a distant
reflector. The observed features can be explained in terms of
absorption if allowance is made for partial covering (i.e. a patchy
absorber).

A partial-covering model was therefore fitted to the data.  The model
comprised a multi-component disc (MCD) model for the soft component,
and a partially covered power-law for the hard continuum. In addition,
this model required a further absorption feature around $\sim 1.2$~keV
(see Section 3.1.3), which is included by subtracting a Gaussian line.
A good fit was obtained ($\chi^2/\nu=510/471$) using two absorbed
power-laws, but a very steep slope, $\Gamma\simeq3.4$ was required,
which appears extremely high given the the present values found in AGN (c.f.
Fig. 1 of Brandt, Mathur \& Elvis 1997). 
Allowing for an exponential cut-off in the power-laws did not
significantly improve the fit ($\chi^2/\nu=509/470$) but gave a more
reasonable spectral slope.  This model is illustrated in Fig.~3.  The
best-fit parameters were as follows: the MCD (colour) temperature $kT
= 0.16$~keV, $\Gamma=2.0$, $\rm E_{\rm cut}=4.4$~keV, and the column
densities (and covering fractions) of the line-of-sight absorbers were
$N_{\rm H} \approx 1.2 \times 10^{22}$ cm$^{-2}$ (0.06) and $ \approx
1.5 \times 10^{23}$ cm$^{-2}$ (0.77). The Fe abundance was about
$10\times$solar with a 90 per cent confidence lower limit of $2 \times
$solar. The average effective slope of the cut-off power-law over
$1-5$~keV is approximately $\Gamma =2.5$.

The temperature and (absorption-corrected) luminosity of the disc
require that its inner radius is only about one gravitational radius
for a $10^6$~M$_{\odot}$ black hole. Assuming the soft X-ray excess
is physically described by a MCD model  then the object would be  super-Eddington
when bright.

\subsubsection{Detection of Fe L absorption}

\begin{figure}
 \includegraphics[width=5.5 cm, angle=270]{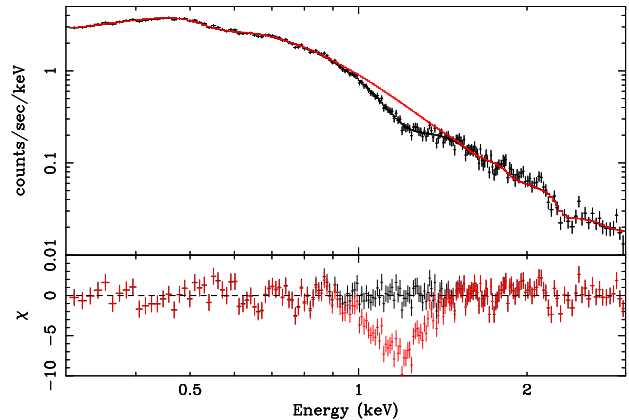}
 \caption{
Spectral fit to the soft energy spectrum obtained with a MCD model
plus partially covered cut-off power-laws. Strong residuals are apparent around 1.2~keV, most
possibly due to Fe-L absorption (light data points)
Including a Gaussian line removes these residuals (dark data points).
The significance of the detection is shown in the lower panel in terms of sigmas.
}
\end{figure}

A broad absorption feature at $E=1.19 \pm 0.02$~keV, with an intrinsic
width $\sigma \approx 0.12$~keV, is required in the above model with
high significance (Fig.~4). The equivalent width is approximately
$EW=120$~eV. The absorbed flux is $1.9_{-0.04}^{+0.02} \times 10^{-13}$
erg cm$^{-2}$ s$^{-1}$. The RGS1 spectrum is consistent with, but does
not constrain, the presence of an absorption feature (no data were
collected from RGS2 in that energy range).  Other features
in the RGS spectra can not be constrained with the available photon statistics.

This deep absorption feature is similar to that reported by Leighly et
al. (1997) from the {\em ASCA} spectrum.  Those authors interpreted
the feature in terms of resonant absorption oxygen from a relativistic
outflow.  A more prosaic explanation was put forward by Nicastro,
Fiore \& Matt (1999) who interpreted the feature as resonance
absorption, primarily by L-shell iron.  Their analysis gave a column
density of the absorbing material of $\log{N_{\rm H}} = 23.5$ and an
ionization parameter of $\log{U} = 1$ with an outflowing and
dispersion velocity of $1000$~km s$^{-1}$ (see also their
Fig.~4). Such an absorption model does not simply explain the large
observed drop at 8~keV.

\subsubsection{An alternative discline explanation}

In the case of 1H~0707--495 (Boller et al. 2002) we noted that the
sharp drop at 7.1~keV could be the blue wing of a strong,
relativistically-blurred iron emission line. This would require some
unusual geometry or situation in order that such a strong line is
produced; the spectrum needs to be reflection-dominated. Fabian et al.
(2002) discuss one possibility involving a corrugated inner accretion
disc; an alternative would be gravitational light bending (e.g.
Martocchia, Matt \& Karas 2002; Fabian \& Vaughan 2003). The spectrum
of IRAS~13224--3809 can also be fitted by such a model (see Fig~5).
Fitting the 2--12~keV spectrum with a power-law continuum plus a
emission line with a Laor (1991) profile worked well with an
emissivity index $q=6.2\pm1.2$, inner and outer radii of $r_{\rm
in}= 1.3^{+0.2}_{-0.1}$ and $r_{\rm out}>180$ gravitational radii for the disc, and an
inclination of $i\approx 60\pm3$~deg. The disc inclination and
emissivity were covariant in the fitting, and a wide range of
inclination angles was possible.  The equivalent width of the line is
unusually high at 5.6~keV. To see whether such a strong line can be
produced, we used a grid of ionized slab models produced by D.
Ballantyne (using the code described by Ross \& Fabian 1993). With the
abundances of metals at 7 times solar and Galactic absorption we
obtain a fit over the entire range from $0.4-12$~keV
with a $\chi_{\nu}^2=1.3$. Some sharp deviations at the 2--3$\sigma$ level
occur at $\sim 6$~keV and again at $\sim 8$~keV. As with 1H~0707--495
this model requires the X-ray spectrum to be reflection-dominated. The
sharp jump in the spectrum at $\sim 1$~keV is due to strong line and
recombination emission by Fe-L, O, and other elements.

\begin{figure}
 \includegraphics[width=5.5 cm, angle=270]{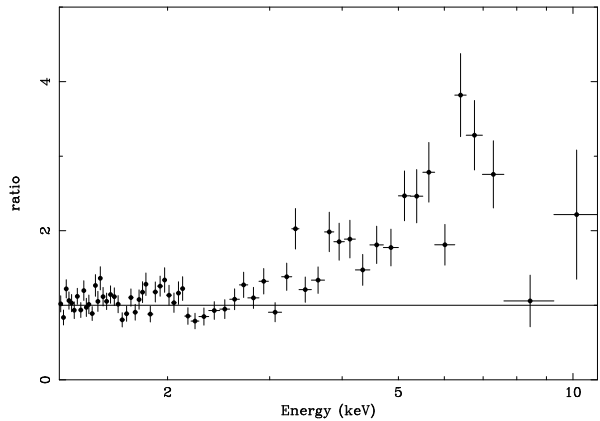}
 \caption{Ratio of the data to a power-law fitted over the 1.3--3~keV
and 8--12~keV bands. The deviations can be seen as a very strong
emission line.}
\end{figure}

\subsection{Flux-dependent spectral properties}

\begin{figure}
 \includegraphics[width=8.0 cm, angle=0]{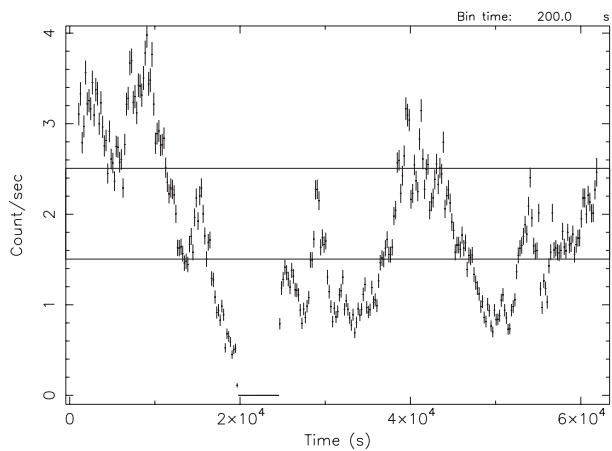}
 \caption{Broad-band ($0.3-12$~keV) pn light curve of IRAS 13224--3809. The bin
size is 200~s. Zero seconds on the time axis marks the start
of the observation at 03:15:02 on 2002-01-19.
The light curve is divided into two high-flux states,
a medium state, and a low state. The marked  interval containing the background flare was omitted.
}
\end{figure}

\begin{figure}
 \includegraphics[width=5.5 cm, angle=270]{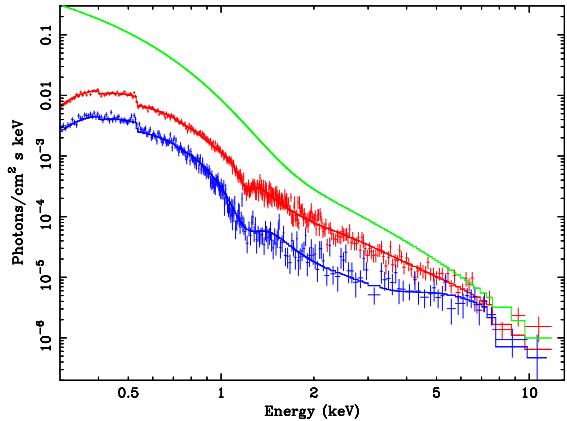}
 \caption{Unfolded spectral fits to the high- and low-flux data (middle and
lower curve, respectively). Note the flattening of the hard
spectrum in the lower flux state. The flattening can be explained
in the partial covering scenario by an increased column density and/or increased
covering fraction of the absorbing material. The upper curve
gives the absorption corrected spectral model.}
\end{figure}

The $0.3-12$~keV light curve of IRAS 13224--3809 shows strong and
rapid variability with flux changes by about a factor of  10 (Fig.~6). We
have divided the light curve into four different flux states: (i) two
high states with count rates above 2.5 counts s$^{-1}$ (time intervals
less than 20 ks and greater than 30 ks after the beginning of the
observations); (ii) a medium state with count rates between 1.5 and 2.5
counts s$^{-1}$; and a (iii) low state below 1.5 counts s$^{-1}$.

Evidence in support of partial covering is obtained from a
comparison of the spectra between a high-flux level and a low-flux
level. The high-flux and low-flux spectra are constructed from
those intervals were the source count rate is $\geq 2.5$ counts
s$^{-1}$ and $\leq 1.5$ counts s$^{-1}$, respectively. The average
fluxes of the two are different by a factor of three. These two
spectra were fitted simultaneously with the same spectral model as
in section~3.1.2 (i.e. a MCD and a cut-off power-law model, and
two absorbed power-law components).  The Fe abundance was fixed at
3$\times$solar. A good fit ($\chi^2/\nu=667/631$) was obtained, as
shown in Fig.~7.  The best fitting parameters were as follows:
$kT=0.16$~keV, $\Gamma=2.0$, $E_{\rm cut}=4.0$~keV. The column
densities (and covering fractions) of the absorbers for the
high-flux spectrum were $N_{\rm H} \approx 1.0$ (0.66) and
$\approx 20$ (0.53), and those for the low-flux spectrum were
$\approx3.1$ (0.53) and $\approx23$ (0.87), where the column is in
units of $10^{22}$ cm$^{-2}$. Thus when the flux dropped the
absorption increased.

\section{Discussion}

IRAS 13224--3809 shows an unusual combination of spectral
features: (i) a sharp spectral drop by a factor of  $\sim5$ at
$\sim 8$~keV, (ii) marginal Fe K emission lines, (iii) a
flattening of the spectrum above 3 keV towards low flux states,
(iv) strong soft X-ray excess emission, (v) a significant Fe-L
absorption feature at around 1.2~keV.

The most plausible explanation for the spectral features (i)
to (iii) appears to be provided by a partial-covering scenario.
The soft X-ray excess is most probably due to a higher
temperature of the accretion disc as often seen in NLS1s. Presently it is not clear
whether the Fe-L absorption feature is physically connected with
the partial covering scenario. The lack of strong emission lines
can be explained if the absorber subtends only a small solid angle
as seen from the central black hole. An immediate consequence,
given the variability in the absorber, is that the absorber must
be located close to the central black hole, maybe even within the
accretion disc region as suggested by e.g Rees (1987), Celotti,
Fabian \& Rees (1992) and Brandt \& Gallagher (2000). The
flattening of the spectrum provides strong evidence for partial
covering, as an increased absorbing column and/or a larger
covering fraction is expected when the source flux is low.
Although the partial coverer model explains the data well, the
present photon statistics above the edge still do not allow us to
constrain precisely the iron abundance, nor can we disentangle the
unique contribution of the covering fraction or the absorbing
column.

The sharpness of the feature, if due to a photoelectric edge, is
surprising. At 8.2~keV the absorbing matter must be partially
ionized (the threshold energies for Fe~{\sc xix--xxiii} are 7.93,
8.07, 8.21, 8.35 and 8.49~keV) in which case a range of ionization
states is expected. These make the observed edge broad rather than
sharp (e.g. Palmeri et al. 2002). Alternatively the feature may be
a neutral edge in approaching matter (say in a wind or jet along
the line of sight, e.g. Chartas et al. 2002). The velocity of the
matter is then required to be $0.15c$. A further possibility if
the absorption is by approaching matter is that the feature is the
start of a trough due to resonance absorption (e.g. Krolik et al.
1985, with Fe~{\sc xxv/xxvi} instead of O~{\sc viii}). There is
then a significant outflow of mass and kinetic energy from the
object.

The extreme spectral properties detected with {\em XMM-Newton}
increase the known spectral complexity in active galactic nuclei
and should further stimulate a combined theoretical and
observational effort to improve our understanding of the physics
of the innermost regions of AGN.  To more precisely constrain the
form of the sharp spectral feature in IRAS 13224--3809 and
distinguish between different models we need a much stronger
signal above 8~keV. This may be possible if the source is observed
during one of its giant amplitude fluctuations.

\section*{Acknowledgments}
Based on observations obtained with XMM-Newton, an ESA science
mission with instruments and contributions directly funded by ESA
Member States and the USA (NASA). WNB acknowledges support from
NASA LTSA grant NAG5-13035.

\bsp
\label{lastpage}
\end{document}